\documentclass{article}
\usepackage{arxiv}

\usepackage{amsmath}
\usepackage{float}
\usepackage{lmodern}
\usepackage{url}
\usepackage{listings}
\usepackage{graphicx}
\usepackage{xcolor}
\usepackage{caption}
\usepackage{natbib}
\usepackage[T1]{fontenc}
\usepackage{makecell}

\usepackage{caption} 

\definecolor{codegray}{gray}{0.95}
\definecolor{commentgray}{gray}{0.4}
\definecolor{arvizblue}{HTML}{2a2eec}
\definecolor{arvizorange}{HTML}{fa7c17}
\definecolor{arvizC4}{HTML}{933708}
\DeclareFontShape{OT1}{cmtt}{bx}{n}{<5><6><7><8><9><10><10.95><12><14.4><17.28><20.74><24.88>cmttb10}{}
\newlength\mystoreparindent

\DeclareUnicodeCharacter{03B2}{$\beta$}
\DeclareUnicodeCharacter{03BC}{$\mu$}
\DeclareUnicodeCharacter{039D}{$\nu$}
\DeclareUnicodeCharacter{03C3}{$\sigma$}
\DeclareUnicodeCharacter{03B8}{$\theta$}

\title{Bambi: A simple interface for fitting Bayesian linear models in Python}

\author{Tomás Capretto \\ IMASL-CONICET \And
        Camen Piho \\ risQ \And
        Ravin Kumar \\ PyMC Labs \And
        Jacob Westfall \\ BlackLocus \And
        Tal Yarkoni \\ University of Texas at Austin \And
        Osvaldo A. Martin \\ IMASL-CONICET \\ Aalto University \\ omarti@unsl.edu.ar\\
}

\begin{document}
\maketitle

\begin{abstract}
The popularity of Bayesian statistical methods has increased dramatically in recent years across many research areas and industrial applications. This is the result of a variety of methodological advances with faster and cheaper hardware as well as the development of new software tools. Here we introduce an open source Python package named Bambi (BAyesian Model Building Interface) that is built on top of the PyMC probabilistic programming framework and the ArviZ package for exploratory analysis of Bayesian models. Bambi makes it easy to specify complex generalized linear hierarchical models using a formula notation similar to those found in R. We demonstrate Bambi's versatility and ease of use with a few examples spanning a range of common statistical models including multiple regression, logistic regression, and mixed-effects modeling with crossed group specific effects. Additionally we discuss how automatic priors are constructed. Finally, we conclude with a discussion of our plans for the future development of Bambi.
\end{abstract}

\section[Introduction]{Introduction}

Bayesian statistics is a flexible and powerful theory that has seen a marked increase in use over the past few years across many classical scientific disciplines like psychology and biology, as well as emerging fields like data science. The primary benefit of using Bayesian methods relative to classical methods is the flexibility when fitting complex and realistic models, incorporation of prior information about plausible values for the model parameters, and the outputs, which are expressed in terms of probabilities which are easier to interpret by non-expert users. However, fitting Bayesian models has historically required considerable mathematical work or large computing resources in order to solve or approximate solutions to difficult statistical problems. While there are still many large or complex Bayesian models that remain computationally challenging, a wide array of useful Bayesian models can now be efficiently fit using average laptops and free, open source, software. Thus, the computational barrier to widespread adoption of Bayesian statistics is rapidly disappearing.

Probabilistic Programming Languages (PPL) have also contributed to minimizing the adoption barrier of Bayesian methods. The aim of a PPL is to augment general purpose programming languages with built-in probabilistic capabilities, allowing applied users to focus on the creation of models, rather than on the implementation of their computation \cite{zotero-null-295,bessiere_bayesian_2013,Ghahramani2015}. While the syntax of PPLs such as PyMC \cite{Salvatier2016} or Stan \cite{Stan2017} is very flexible, it could still be too verbose for many practitioners, such as those coming from a frequentist paradigm, the R programming language, or other Python packages like statsmodels \cite{Pedregosa2011} or Scikit-learn \cite{Seabold2010}. 

Even for users that are experts in PPLs it still takes time to write a simple linear model. These users can benefit from a more compact syntax closer to the formula notation found in popular R packages for statistical computing; like lme4 \cite{lme4}, rstanarm \cite{Goodrich2020} or brms \cite{brms}.  The development of a simpler, more intuitive interface would go a long way towards promoting the widespread adoption of Bayesian methods within applied scientific fields as well as industrial applications.

Here we introduce Bambi (BAyesian Model Building Interface) an open source Python package designed to make it considerably easier for practitioners to fit Generalized Linear Multilevel Models (GLMMs)\footnote{Also known as generalized mixed linear models} using a Bayesian approach. Generalized linear multilevel models encompass a large class of techniques that include most of the models commonly used in applied fields of research: linear regression, ANOVA, logistic and Poisson regression, multilevel or hierarchical modeling, crossed group specific effects models. Bambi is built on top of the PyMC Python package, which implements a state of the art adaptive dynamic Hamiltonian Monte Carlo algorithm \cite{Hoffman2014, Betancourt2017}, among other sampling methods. Bambi also uses ArviZ \cite{Kumar2019} for comprehensive sampling diagnostics, model criticism, model comparison and visualization of Bayesian models. Importantly, Bambi affords both ease of use and considerable power: for example, beginner users can quickly specify complex generalized linear multilevel models with sensible default priors and syntax similar to popular R statistical computing packages, while advanced users can still directly access all the internal objects exposed by PyMC, allowing strong flexibility as well as a sensible learning progression.

The remainder of this article will focus primarily on Bambi in practice. We start by giving a brief overview of the family of models supported in Section \ref{sec:overview}, then we continue with Section \ref{examples} demonstrating basic usage through examples, Section \ref{sec:prior_choice} discussing default prior choices, and Section \ref{sec:formula} providing further insights into the inner workings of the formula specification. Finally, we conclude with Section \ref{discussion} discussing the limitations and the future of the Bambi package.

Bambi is available from the Python Package Index at \url{https://pypi.org/project/bambi/}. Alternatively, it can be installed using conda. The project is hosted and developed at \url{https://github.com/bambinos/bambi}. The package documentation, including installation instructions and many examples of how to use Bambi to conduct different statistical analysis, can be found at \url{https://bambinos.github.io/bambi/}.

The version of Bambi used for this article is 0.6.3. All analyses are supported by extensive documentation in the form of interactive Jupyter notebooks \cite{Kluyver2016} available in the paper repository on GitHub \url{http://github.com/bambinos/paper}, enabling readers to re-run, modify, and otherwise experiment with the models described here on their own machines. This repository also includes instructions on how to set up an environment with all the dependencies used when writing this manuscript.

\section{A model overview}
\label{sec:overview}

This section provides a brief review of GLMMs, the family of models supported by Bambi, and introduces the \texttt{Model} class, which is the class we use to create models in Bambi.

Generalized Linear Models (GLMs) extend standard linear regression models to include non-Gaussian response distributions and nonlinear functions of the mean. GLMMs then extend GLMs to incorporate group specific effects (also known as random effects in frequentist literature). This allows the modeling of hierarchically structured data, taking complex dependency structures into account.

A generalized linear multilevel model is defined as

\begin{equation}
    Y_i \sim \mathcal{D}(g^{-1}(\eta_i), \boldsymbol{\theta})
\end{equation}

In this equation, $Y_i$ represents the random variable we want to model, $\mathcal{D}$ represents the distribution family parameterized by the mean, and $\boldsymbol{\theta}$ represents auxiliary parameters specific to each family such as the standard deviation $\sigma$ in Gaussian models or the shape parameter $\alpha$ in Gamma models. $g$ is a differentiable and monotonic function known as the link function, and $\eta_i$ is a linear combination of the predictors known as the linear predictor. 

We can write the linear predictor for all the observations as 

\begin{equation}
    \boldsymbol{\eta} = \mathbf{X}\boldsymbol{\beta} + \mathbf{Z}\mathbf{u}
\end{equation}

where $\boldsymbol{\beta}$ and $\mathbf{u}$ are the coefficients for the common effects and group specific effects respectively and, $\mathbf{X}$ and $\mathbf{Z}$ are their corresponding design matrices.
In contrast to the frequentist paradigm, where $\boldsymbol{\beta}$ is treated as an unknown fixed vector of coefficients and $\mathbf{u}$ is treated as an unobserved random effect that is part of the error term, we treat both vectors of coefficients $\boldsymbol{\beta}$ and $\mathbf{u}$ as random variables whose joint posterior distribution is estimated from the data.

All models in Bambi are instances of the \texttt{Model} class. These objects contain all the methods we use to specify, fit, and analyze Bayesian models. We now briefly discuss the arguments most relevant for most users;  \texttt{formula}, \texttt{data}, \texttt{family}, \texttt{priors}, and \texttt{link}. 

\begin{itemize}
    \item \texttt{formula}: A string that describes the model we want to fit using a formula notation very close to the one in the R language. As in the R library lme4, the pipe operator \texttt{|} is used to specify group specific terms.  A more detailed review of formulas is given in Section \ref{sec:formula}.
    \item \texttt{data}: The dataset where variables in the model are taken from. This is a pandas \cite{mckinney-proc-scipy-2010} \texttt{DataFrame}, or the path to a CSV file that can be read as such. 
    \item \texttt{family}: The model family, analogous to families in R. This can be a string with the name of a built-in family (see Table \ref{table:family_table}) or a custom family created with the \texttt{Family} class. The default is \texttt{"gaussian"}.
    \item \texttt{priors}: An optional dictionary containing specifications for prior distributions for one or more terms in a model. If no priors are specified, Bambi will choose priors for the parameters according to the methods in Section \ref{sec:prior_choice}. Section \ref{sec:hierarchical_models} contains an example on how to use this argument to specify custom priors.
    \item \texttt{link}: The name of the link function to use. Bambi comes with several built-in link functions. If the user does not specify a link function, it is automatically determined by Bambi according to the \texttt{family} (See Table \ref{table:family_table}).
\end{itemize}

\section{Usage examples}
\label{examples}

In this section, we provide a high-level overview of the Bambi package and the supported models. We illustrate its use via a series of increasingly complex applications, beginning with a straightforward multiple regression model, and culminating with a linear mixed model that involves custom priors.

\subsection{Multiple linear regression}

We begin with an example from personality psychology. The data that we consider come from the Eugene-Springfield community sample \cite{Goldberg1999}, a longitudinal study of hundreds of adults who completed dozens of different self-reports and behavioral measures over the course of 15 years. Among the behavioral measures is a numerical index of illegal drug use (the “drugs” outcome from the Behavioral Report Inventory; for details, see \cite{Grucza2007}). We wish to know: which personality traits are associated with higher or lower levels of drug use? In particular, how do participants' standings on the “Big Five” personality dimensions predict drug use? The “Big Five” personality dimensions are Openness to experience (o), Conscientiousness (c), Extraversion (e), Agreeableness (a), and Neuroticism (n). This dataset can be loaded as a pandas \texttt{DataFrame} using the function \texttt{load\_data()} from Bambi. Then it is simple to specify a multiple regression model using a formula-like interface:

\begin{lstlisting}
# Load the Bambi library
import bambi as bmb
# Load a predefined dataset
data = bmb.load_data("ESCS")
# Define the model
model = bmb.Model("drugs ~ o + c + e + a + n", data)
# Fit model
idata = model.fit(draws=2000)
\end{lstlisting}

This fully specifies a Bambi model and fits it. By default, Bambi uses a Gaussian response and the identity link function. Notice that we have not specified prior distributions for any of the parameters. When no priors are explicitly specified by the user, Bambi will choose default priors for the parameters of the model. See Section \ref{sec:prior_choice} for details. We can obtain a summary of the model together with the priors by printing the \texttt{model} object

\begin{lstlisting}
Formula: drugs ~ o + c + e + a + n
Family name: Gaussian
Link: identity
Observations: 604
Priors:
  Common-level effects
    Intercept ~ Normal(mu: 2.2101, sigma: 21.1938)
    o ~ Normal(mu: 0.0, sigma: 0.0768)
    c ~ Normal(mu: 0.0, sigma: 0.0868)
    e ~ Normal(mu: 0.0, sigma: 0.0816)
    a ~ Normal(mu: 0.0, sigma: 0.0973)
    n ~ Normal(mu: 0.0, sigma: 0.0699)

  Auxiliary parameters
    sigma ~ HalfStudentT(nu: 4, sigma: 0.6482)
\end{lstlisting}

\begin{figure}[htb]
\centering
\includegraphics[width=5.5in]{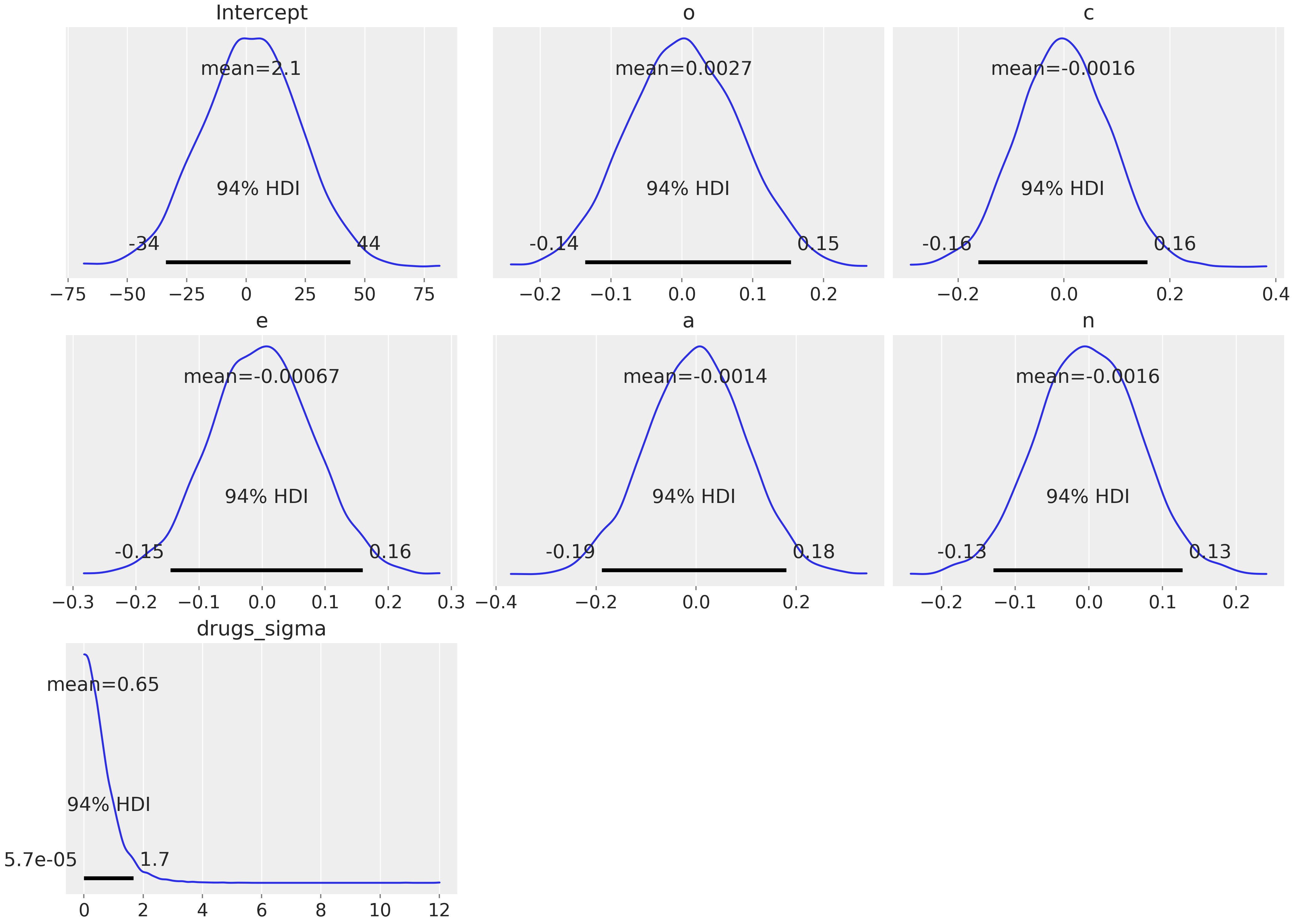}
\caption[]{Density estimates based on 5000 samples from the prior distribution for all the regression coefficients. If the user does not explicitly state the priors to be used for the model parameters, Bambi will choose default prior distributions sensible in a wide range of use cases.}
\label{fig:drugs_priors}
\end{figure}

We can also inspect the priors visually using the command:

\begin{lstlisting}[numbers=none]
model.plot_priors()
\end{lstlisting}

This will return a figure similar to Figure \ref{fig:drugs_priors}, which shows estimates for the prior distributions based on computationally simulated draws. While not as mathematically precise as closed-formed expressions, the use of simulation removes mathematical limitations by allowing the user to explore complex priors specifications and compute posteriors that might not have closed-form solutions.

Notice that the standard deviations of the priors for the slopes seem to be quite small, with most of the probability density being between -.15 and .15. This is due to the relative scales of the outcome and the predictors: the outcome, \texttt{drugs}, is a mean score that ranges from 1 to about 4, while the predictors are all sum scores that range from about 20 to 180. So a one unit change in any of the predictors, which is a trivial increase on their scale, is likely to lead to a very small absolute change in the outcome. These priors are actually quite wide (or \textit{weakly informative}).

By default, Bambi fits models using an adaptive dynamic Hamiltonian Monte Carlo algorithm, \cite{Hoffman2014, Betancourt2017} which samples from the joint posterior distribution of the parameters. Bambi will attempt to use the available number of CPUs cores in the system to run between 2 and 4 chains in parallel. Running more than one chain is useful to run inference diagnostics (see Table \ref{table:summary}). 

In the above example, the optional argument \texttt{draws} indicates that we want to obtain 2000 draws, per chain, from the posterior distribution. Bambi will also run a certain number of iterations to tune the sampling algorithm (defaults to 1000). These tuning draws will be discarded by default, as they are not valid draws from the posterior distribution.

\begin{figure}[htb]
\centering
\includegraphics[width=5.5in, height=4.5in]{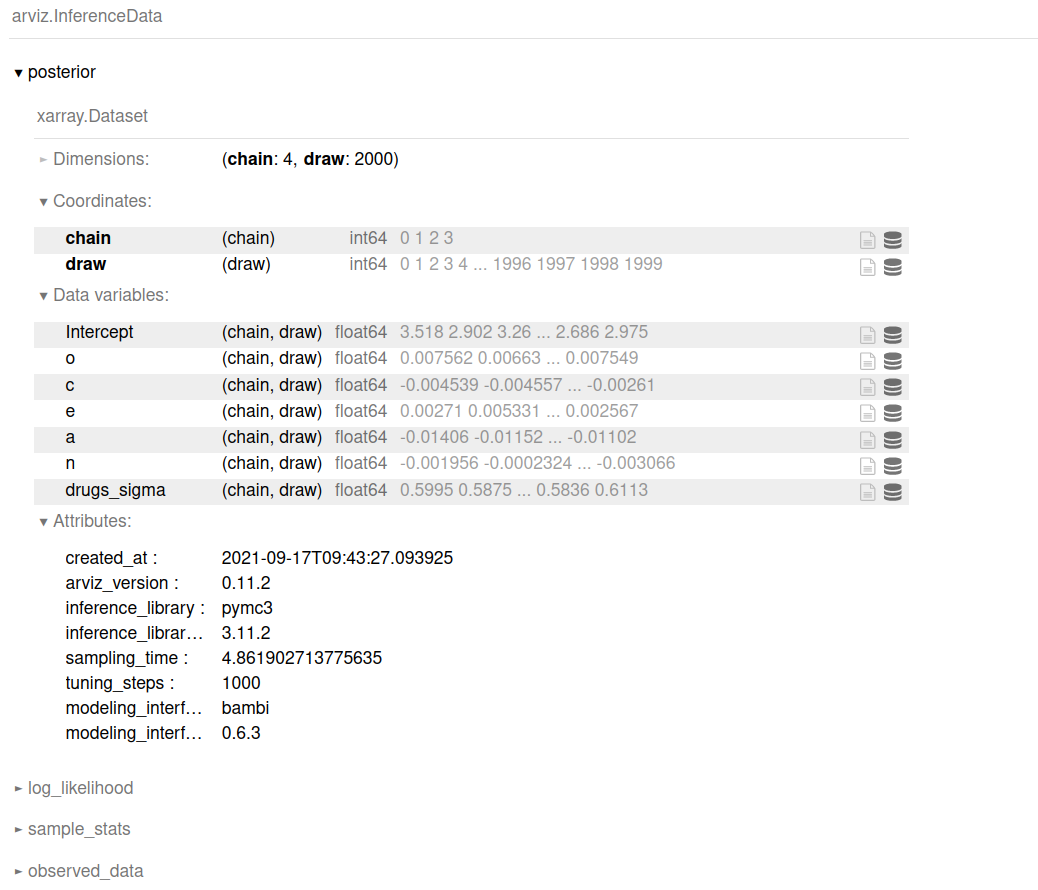}
\caption[]{HTML representation of an \texttt{InferenceData} object. We can see information is stored into four groups: \texttt{posterior}, \texttt{log\_likelihood}, \texttt{sample\_stats}, and \texttt{observed\_data.} Other groups not shown here are also possible. The posterior group is unfolded showing information like the Dimensions (4 chains with 2000 draws each), and the Data variables including the coefficients for the predictors \texttt{o}, \texttt{c}, \texttt{e}, \texttt{a}, \texttt{n} we explicitly added when defining the model plus an \texttt{Intercept} and the standard deviation of the Gaussian likelihood \texttt{drugs\_sigma}. Metadata information like which libraries' version was used and date of creation of files are also available.}
\label{fig:InferenceData}
\end{figure}

Once the posterior sampling finishes, the result is saved into an \texttt{InferenceData} object, like the \texttt{idata} object in the above example. Such objects contain data related to the model divided into groups, like \texttt{posterior}, \texttt{observed\_data}, \texttt{posterior\_predictive}, etc., as can be seen in Figure \ref{fig:InferenceData}. The \texttt{InferenceData} objects can be passed to many functions in ArviZ to obtain numerical and visual diagnostics, and plots in general. For example, with the command \texttt{az.summary(idata)} we can get a summary of the posterior (including the mean, standard deviation, and Highest Density Intervals) and also diagnostics of the sampling, (including the Monte Carlo standard error, effective size samples and $\hat R$ \cite{Vehtari_2020}). Table \ref{table:summary} shows an example. 

\begin{table}[t!]
\centering
\scriptsize
\begin{tabular}{llllllllllllp{7.4cm}}
\hline
{} &   mean &     sd &  hdi\_3\% &  hdi\_97\% &  mcse\_mean &  mcse\_sd &  ess\_bulk &  ess\_tail &  r\_hat \\
Intercept   &  3.307 &  0.361 &   2.631 &    3.983 &      0.004 &    0.003 &    7839.0 &    6576.0 &    1.0 \\
o           &  0.006 &  0.001 &   0.004 &    0.008 &      0.000 &    0.000 &    7210.0 &    6782.0 &    1.0 \\
c           & -0.004 &  0.001 &  -0.007 &   -0.001 &      0.000 &    0.000 &    7555.0 &    6833.0 &    1.0 \\
e           &  0.003 &  0.001 &   0.001 &    0.006 &      0.000 &    0.000 &    7806.0 &    7103.0 &    1.0 \\
a           & -0.012 &  0.001 &  -0.015 &   -0.010 &      0.000 &    0.000 &    9261.0 &    6844.0 &    1.0 \\
n           & -0.002 &  0.001 &  -0.004 &    0.001 &      0.000 &    0.000 &    8087.0 &    6908.0 &    1.0 \\
drugs\_sigma &  0.592 &  0.017 &   0.559 &    0.623 &      0.000 &    0.000 &    8985.0 &    6235.0 &    1.0 \\
\end{tabular}
\caption{Numerical summary of the posterior and sample diagnostics. The first four columns, mean, sd (standard deviation) hdi\_3\% and hdi\_97\% (the boundaries of the highest density intervals 94\%) provide posterior summaries. The rest of the columns are sample diagnostics that could help us to assess the quality of the approximated posterior. mcse\_mean and mcse\_sd, are the Monte Carlo standard error for the mean and standard deviation, respectively. The ess represents the effective sample size, the closest this number to the total number of draws (chains x draws) the better, and it should not be lower than (chains x 50). The ess is not the same for different regions of the parameter space, thus here we show it for the bulk of the distribution and tails. r\_hat is the $\hat R$ diagnostic. Ideally, this number should be smaller than 1.01. Larger numbers indicate convergence issues of the sampling method \cite{Vehtari_2020}.}
\label{table:summary}
\end{table}

\begin{figure}[!htb]
\centering
\includegraphics[width=6in, height=7in]{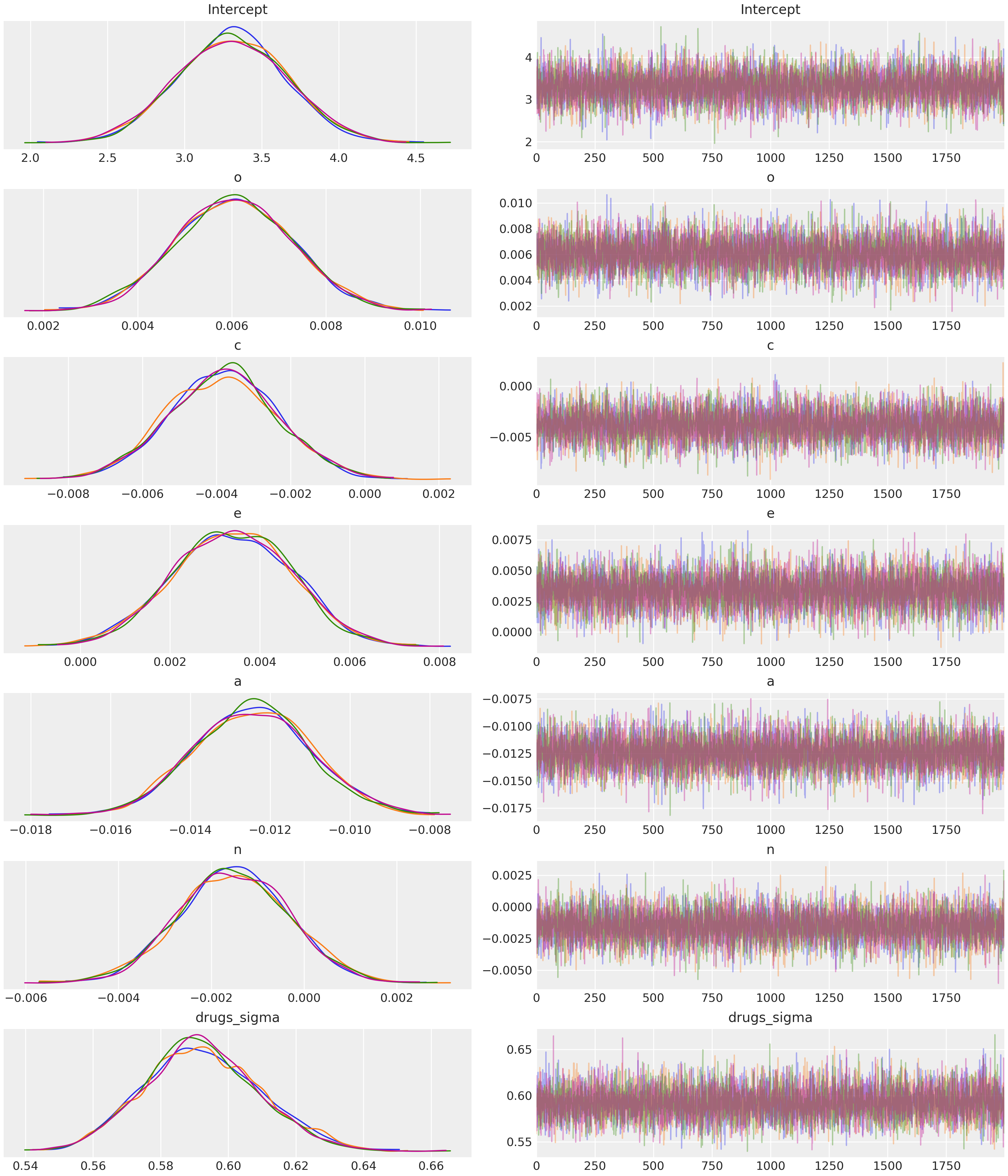}
\caption[]{The left panels show the kernel density estimates for the marginal posterior distributions for all the model’s parameters, which summarize the most plausible values of the regression coefficients, given the data we have observed. These posterior density plots have four overlaid distributions because we ran four chains in parallel. The panels on the right are “trace plots” showing the sampling paths of the four chains as they wander through the parameter space. If any of these paths exhibited high autocorrelation we would be concerned about the convergence of the chains.}
\label{fig:drugs_posterior}
\end{figure}

A common way to visually explore the posterior is with the command \texttt{az.plot\_trace(idata)}. This command results in Figure \ref{fig:drugs_posterior}. The left panels show the kernel density estimates of the marginal posterior distributions for all the model's parameters, i.e., the probability distribution of the plausible values of the regression coefficients, given the model and data we have observed. These posterior density plots show four overlaid distributions because we run four chains. The right panels of Figure \ref{fig:drugs_posterior} show the sampling paths (or \textit{traces}) of the four chains as they wander through the parameter space, this is after tuning draws were discarded. These \textit{trace plots} are useful for sampling diagnostic purposes \cite{Raftery1996}. In this example, the traces mix well and show a stationary pattern. If any of these paths did not mix well or showed a trend, we would be concerned about convergence, and we should work to fix the convergence issues before continuing with other analysis \cite{BayesianWorkflow2020, martin2021}.

From the left-panel in the Figure \ref{fig:drugs_posterior} we can see the model results suggest that 4 of the 5 personality dimensions, all but Neuroticism (n), have at least some non-trivial association with drug use. According to the sign of their coefficients, we can conclude that higher scores for Openness (o) and Extraversion (e) are associated with a higher drug use index, while higher scores for Consciousness (c) and Agreeableness (a) are related to lower values of the drug use index. Finally, since the marginal posterior of Neuroticism (n) has a non-negligible probability around zero, we conclude it is not associated with drug use if we are already taking into account all the other variables in the model.

We may further be interested in asking: which of these personality dimensions matter more for the prediction of drug use? There are many possible ways to think about what it means for one predictor to be relatively “more important” than another predictor \cite{Hunsley2003, Westfall2016}, but one conceptually straightforward way to approach the issue is to compare partial correlations between each predictor with the outcome, controlling for all the other predictors. These comparisons are somewhat challenging using traditional frequentist methods, perhaps requiring a bootstrapping approach, but they can be formulated very naturally in the Bayesian framework thanks to Bambi and the libraries it relies on.  We can simply apply the relevant transformation to all the posterior samples to obtain the joint posterior distribution on the (squared) partial correlation scale. 

\begin{figure}[htb]
\centering
\includegraphics[width=5.5in]{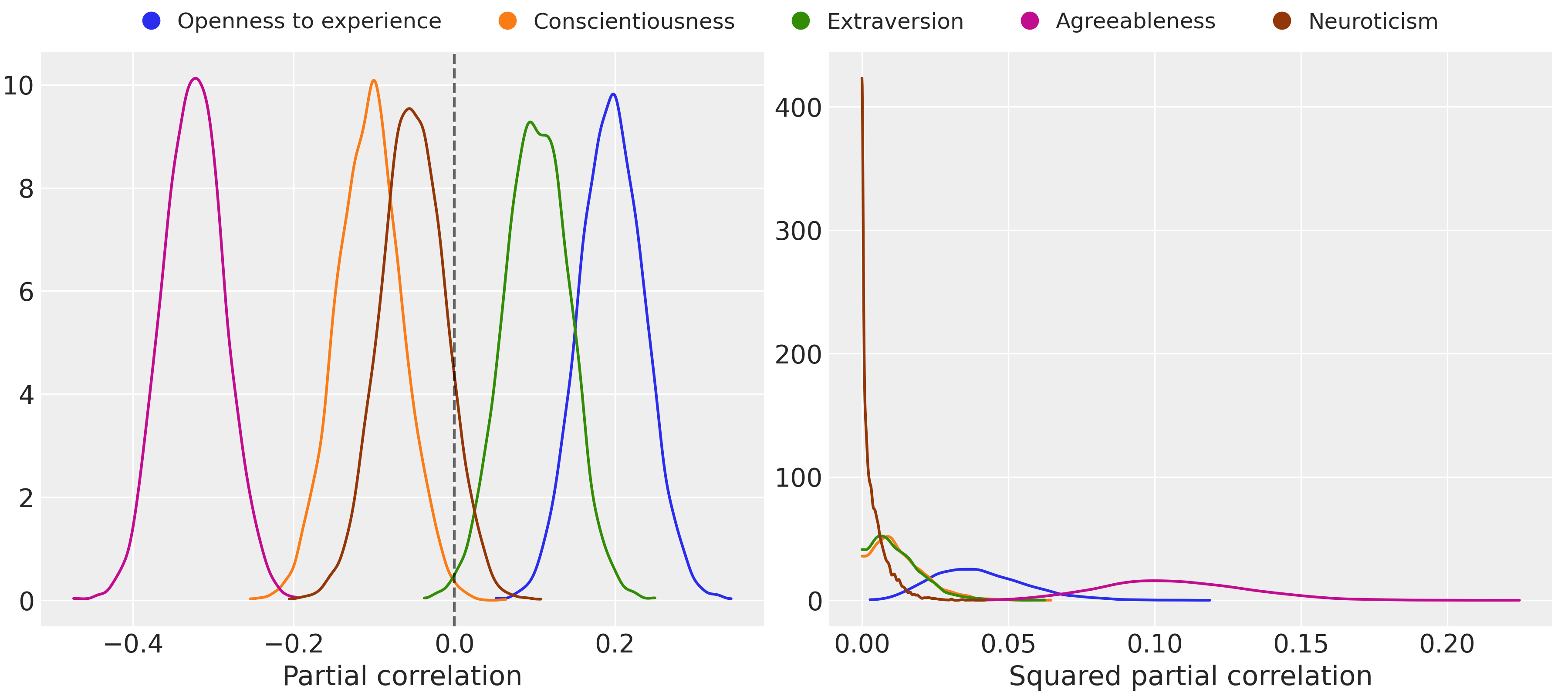}
\caption[]{Posterior distributions of the relationships between the Big Five predictors and drug use on the
partial correlation (left) and squared partial correlation scales (right).}
\label{fig:drugs_partial_correlations}
\end{figure}

Pearson's partial correlation is a measure of linear association between a predictor and the outcome after controlling for the set of all other predictors in the model. In plain English, the partial correlation of a given predictor is a measure of how much information about the outcome is explained by that predictor itself, and not by the others. It is possible to convert the regression coefficient into a partial correlation by multiplying it by a constant that depends on the variability of the predictor and the outcome, and the degree of linear association with the set of other predictors. The derivation of this term, together with the code we used, is included in the Appendix \ref{sec:multiple_regression_appendix} as well as in the online notebooks. That is what we have done to the slope samples before obtaining Figure \ref{fig:drugs_partial_correlations}. These marginal posteriors allow us to visualize the plausible values for the partial correlations and squared partial correlation and quickly see, for example, that there is a negative association between Agreeableness (a) and the drugs index or that Openness (o) contains more information about drug usage than Extroversion (e) because the marginal posterior of the former is concentrated around larger values on the right panel.

\begin{figure}[htb]
\centering
\includegraphics[width=5in]{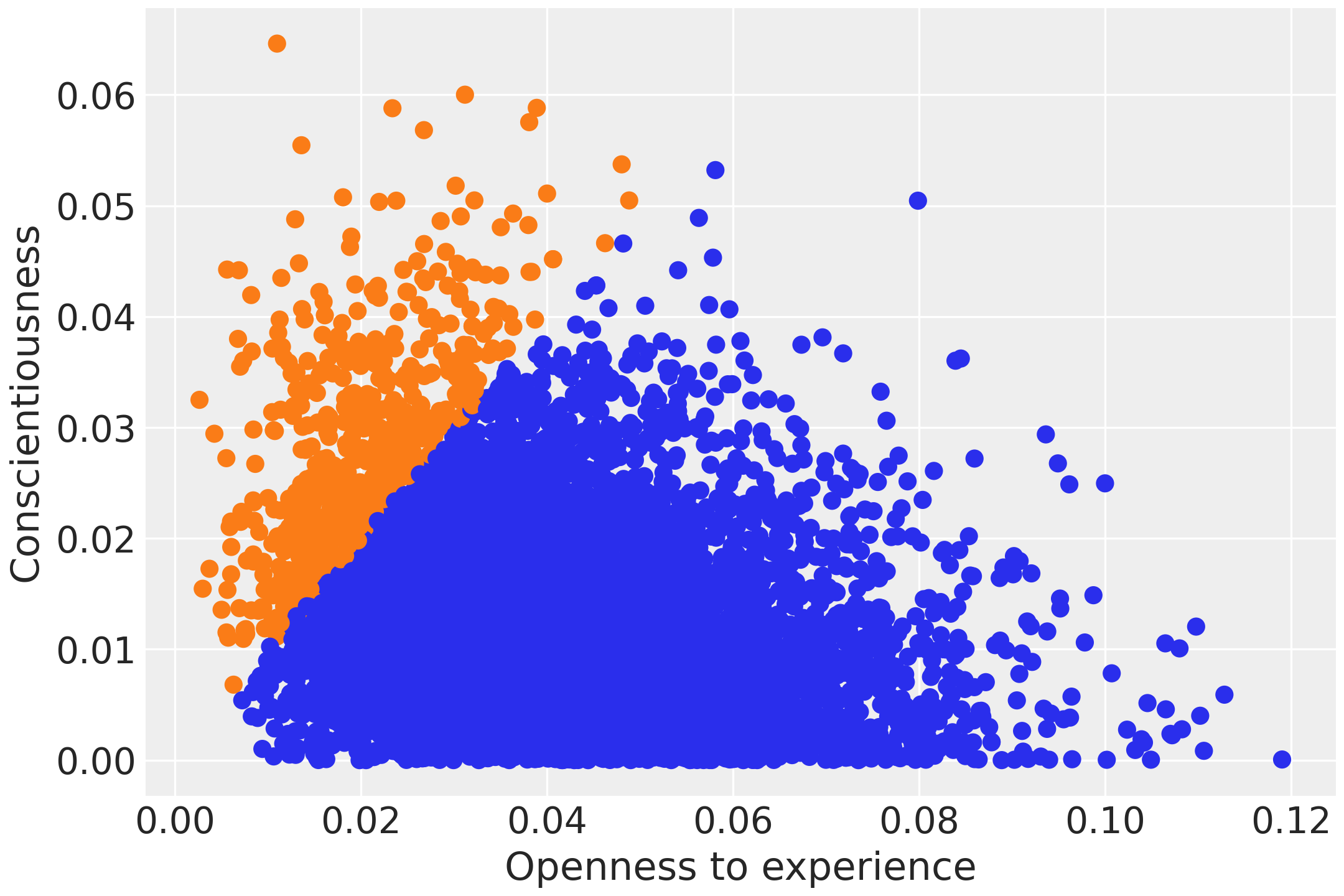}
\caption[]{Draws from the 2-dimensional joint posterior of openness and conscientiousness in terms of squared partial correlation. Orange dots represent draws where conscientiousness has a larger squared partial correlation than openness, and blue dots represent draws where openness is the one with larger values. The percentage of blue dots (93\%) represents the probability that openness is a stronger predictor than conscientiousness.}
\label{fig:drugs_scatter}
\end{figure}

We can also use the joint posterior to draw conclusions about questions involving the partial correlation of more than one predictor. For example, we can conclude that the probability that Openness to experience (o) is a stronger predictor than Conscientiousness (c) is about 93\% (Figure \ref{fig:drugs_scatter}) or that the probability that Agreeableness is the strongest of the five predictors is about 99\% (No figure is shown, as this involves a 5-dimensional posterior).

\begin{figure}[h!tb]
\centering
\includegraphics[width=4in]{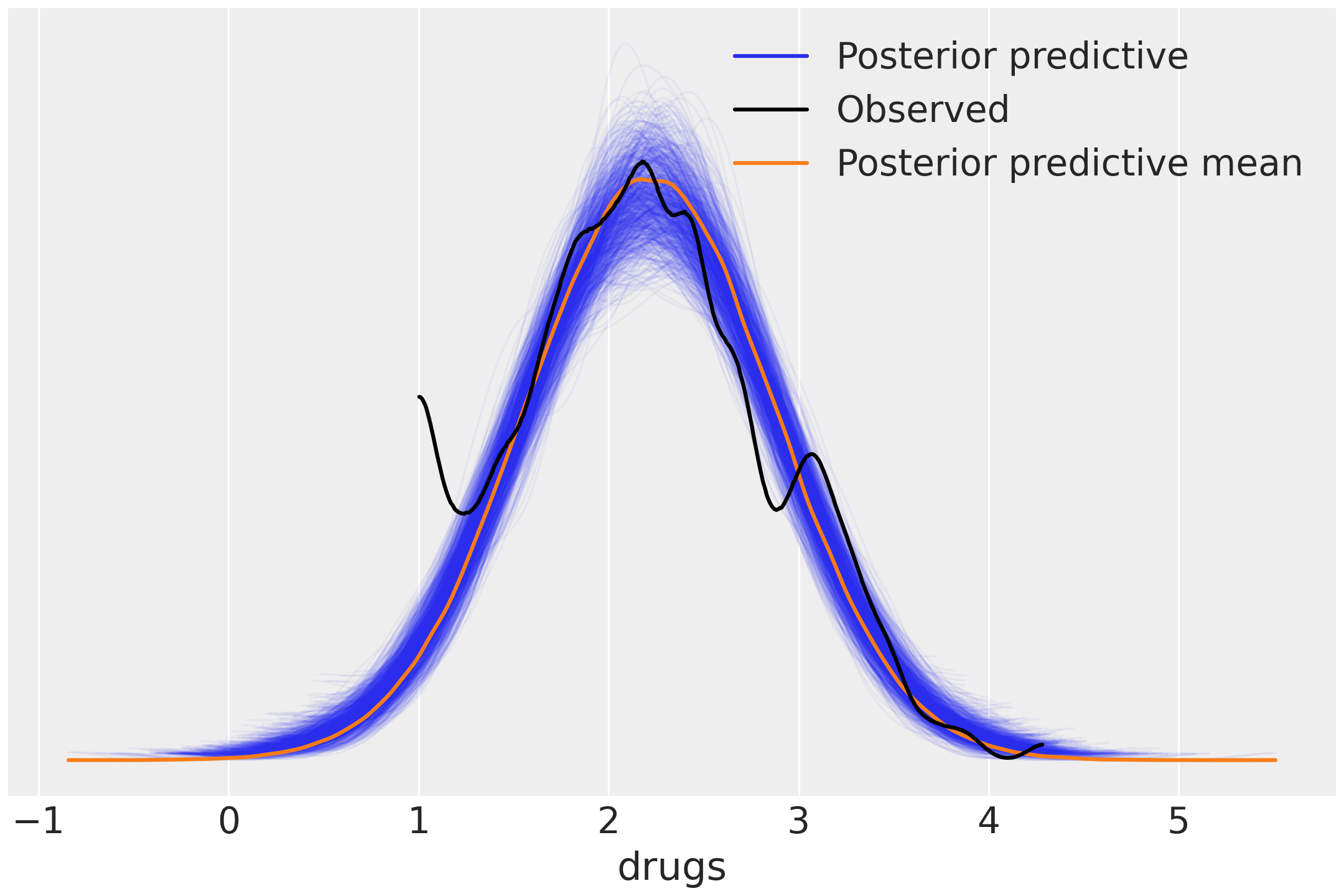}
\caption[]{Posterior predictive plot of Big Five personality dimensions. The blue lines represent samples from the posterior predictive distribution, and the black line represents the observed data. The posterior predictions seem to adequately represents the observed data in all regions except near the value of 1, where the observed data and predictions diverge.}
\label{fig:BigFivePPC}
\end{figure}

We may also use the posterior distribution to compute the posterior predictive distribution. As the name implies, these are predictions assuming the model's parameters are distributed per the estimated posterior. Thus, the posterior predictive includes the uncertainty about the parameters. The posterior predictive evaluated at the observed values of the predictors is a common method to diagnose a model's fit, and can also be used to compare two or more models. Using Bambi, the posterior predictive distribution, evaluated at the observed values of the predictors, can be computed using \texttt{model.predict(idata, kind="pps")}. This \texttt{.predict()} method, which we discuss more deeply in the following example, automatically adds the posterior predictive sample to the \texttt{InferenceData} object \texttt{idata}. We can then use ArviZ to plot Figure \ref{fig:BigFivePPC}.

\begin{lstlisting}
model.predict(idata, kind="pps")
az.plot_ppc(idata)
\end{lstlisting}

\subsection{Logistic regression}

Our next example involves an analysis of the American National Election Studies (ANES) data. The ANES is a nationally representative, cross-sectional survey used extensively in political science. We will use a dataset from the 2016 pilot study, consisting of responses from 1200 voting-age U.S. citizens, from  \url{http://electionstudies.org}. From this dataset, we extracted the subset of 421 respondents who had observed values on the following variables:

\begin{itemize}
    \item \texttt{vote}: If the 2016 presidential election were between Hillary Clinton for the Democrats and Donald Trump for the Republicans, would the respondent vote for Hillary Clinton, Donald Trump, someone else, or probably not vote? Observed answers are coded as \texttt{"clinton"}, \texttt{"trump"}, and \texttt{"someone\_else"}.
    \item \texttt{party\_id}: With which US political party does the respondent usually identify? For example, Republican, Democrat, or Independent. Observed answers are \texttt{"democrat"}, \texttt{"republican"}, and \texttt{"independent"}.
    \item \texttt{age}: Computed from the respondent's birth year. This is a numerical variable that ranges from 18 to 95.
\end{itemize}

For brevity of presentation, we focus only on data from respondents who indicated that they would vote for either Clinton or Trump, and we will model the probability of voting for Clinton.

As expected, respondents who self-identify as Democrats are more likely to say they would vote for Clinton over Trump; respondents who self-identify as Republicans report an intention to vote for Trump over Clinton; and Independent respondents fall somewhere in between. What we are interested in is the relationship between respondent age and intentions to vote for Clinton, and in particular, how age may interact with party identification in predicting voting intentions. 

As before, we load the data as a pandas \texttt{DataFrame} using the \texttt{bmb.load\_data()} function. Then we can specify and fit the logistic regression model using the following commands: 

\vspace{10pt}
\begin{minipage}{\linewidth}
\begin{lstlisting}
# Load data
data = bmb.load_data("ANES")
# Keep only records where vote goes to Clinton or Trump
data = data.loc[data["vote"].isin(["clinton", "trump"]), :]
# Create model
model = bmb.Model("vote[clinton] ~ party_id + party_id:scale(age)", data, 
                  family="bernoulli")
# Fit model
idata = model.fit(draws=2000, tune=2000)
\end{lstlisting}
\end{minipage}
\vspace{10pt}

We name the model \texttt{model} and we use \texttt{vote[clinton]} to tell Bambi that we are modeling the probability of voting for Clinton. The latter is optional syntax that we use on the left-hand-side of the formula to explicitly ask Bambi to model the probability the variable \texttt{vote} is equal to \texttt{clinton} (case-sensitive). If unlike \texttt{clinton}, the response we would like to model had spaces, we would have had to wrap it within single quotes, for example, \texttt{vote[\textquotesingle hillary clinton\textquotesingle]}. Note however this step is not strictly necessary, as Bambi will pick a reference category and include it in the output if we do not pass one explicitly. Another option is to encode \texttt{vote} as a \texttt{0-1} variable before creating the model and Bambi will model the probability the variable is equal to \texttt{1}. We set \texttt{family="bernoulli"} because the outcome variable, \texttt{vote}, represents Bernoulli trials, where \texttt{vote=="clinton"} represents a success and \texttt{vote=="trump"} represents a failure. We could have also specified \texttt{link="logit"} to indicate the link function of the GLM, but the logit link function is the default when \texttt{family="bernoulli"} (see Table \ref{table:family_table}). As before, we instruct Bambi to sample 2000 draws from the joint posterior, but now we also ask for 2000 tune steps.

We can also see the variable \texttt{age} is wrapped with the function \texttt{scale()}. This function standardizes the variable before entering the model by subtracting its mean and dividing by its standard deviation. As a result, we get roughly a 2x speedup in the sampling process for this particular example.

\texttt{scale()} is one of the built-in stateful transformations provided by formulae \cite{bambinoscapretto2021}, the library Bambi uses it to translate model formulas into design matrices. These transformations are stateful because they remember the state of the original dataset, and use it in transforming new datasets. For example, the \texttt{scale()} transformation remembers the mean and standard deviation of the variable \texttt{age} in the original dataset \texttt{data}. These values are used later when making predictions for a new dataset to standardize the new \texttt{age} values with the original mean and standard deviation.

\begin{figure}[!htb]
\centering
\includegraphics[width=6in, height=7in]{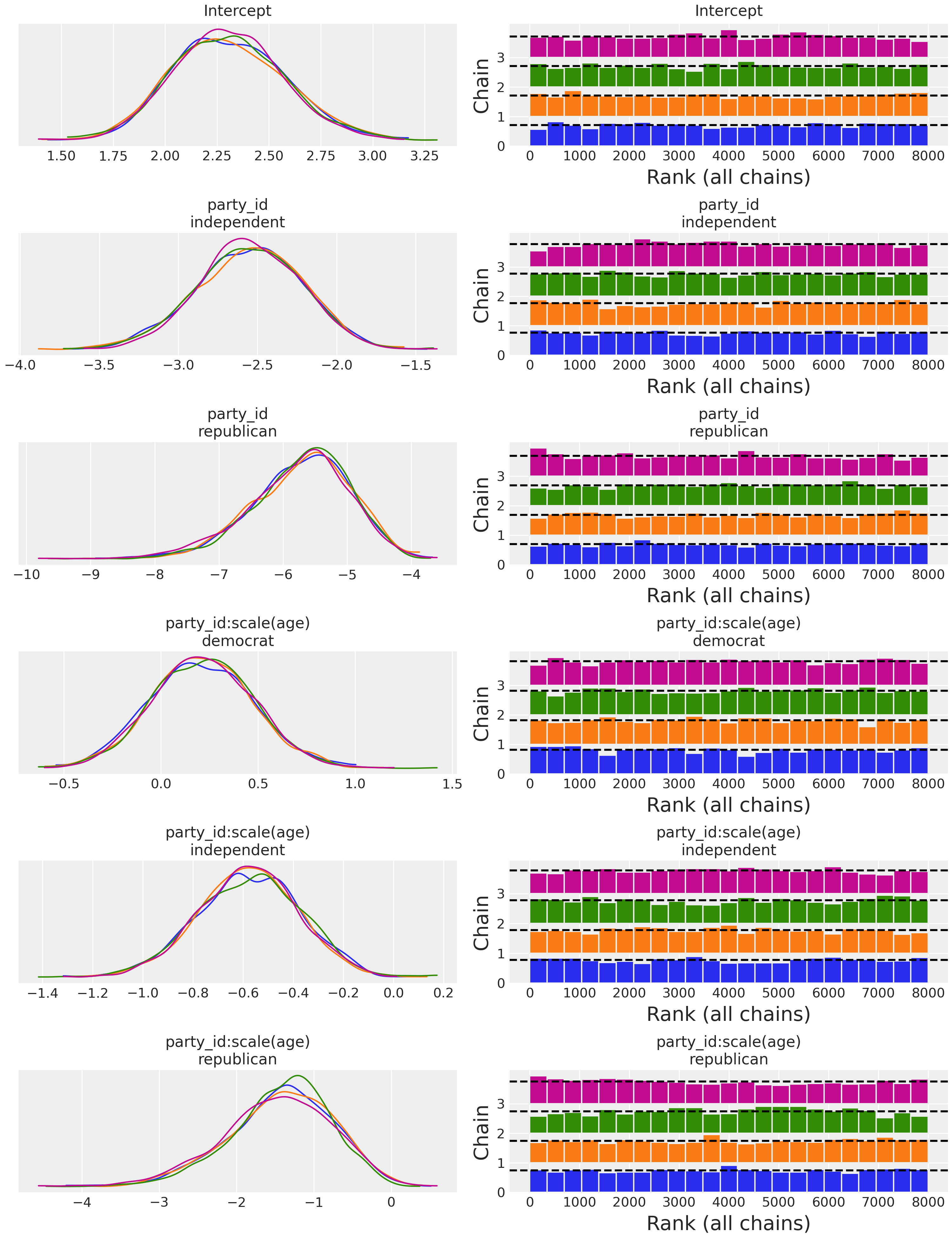}
\caption[]{The left panels show the kernel density estimates for the marginal posterior distributions for all the model’s parameters. The right panels show rank plots. This type of visualization is based on histograms of the ranked posterior draws (ranked over all chains) plotted separately for each chain. If all the chains are targeting the same posterior, we expect the ranks in each chain to be uniform, whereas if one chain has a different location or scale parameter, this will be reflected in the deviation from uniformity \cite{Vehtari_2020}. The similarity of the rank plots in this example indicates the chains mixed well.} 
\label{fig:clinton_posterior}
\end{figure}

Again, we can use the command \texttt{az.plot\_trace(idata)} to obtain plots for the marginal posteriors. Figure \ref{fig:clinton_posterior} shows the output of \texttt{az.plot\_trace(idata, kind="rank\_bars")}. The argument \texttt{kind="rank\_bars"} indicates to use a rank plot based on bars instead of the classic trace plot to explore the mixing and convergence of the chains.

\begin{table}[t!]
\centering
\begin{tabular}{lllp{7.4cm}}
\hline
Family & Response & Available links \\  \hline
bernoulli & Bernoulli & \textbf{logit}, probit, cloglog, identity \\
beta & Beta & \textbf{logit}, probit, cloglog, identity \\
binomial & Binomial & \textbf{logit}, probit, cloglog, identity \\
gamma & Gamma & \textbf{inverse}, identity, log \\
gaussian & Normal & \textbf{identity}, log, inverse \\
negativebinomial & NegativeBinomial & \textbf{log}, identity, cloglog \\
poisson & Poisson & \textbf{log}, identity \\
t & StudentT & \textbf{identity}, log, inverse \\
wald & InverseGaussian & \textbf{inverse\_squared}, inverse, identity, log
\end{tabular}
\caption{Summary of the currently available families and the link functions they can use with their defaults in bold. These families are represented in Bambi with a \texttt{Family} class. This class contains the conditional distribution of the response (as the name of a valid PyMC distribution), the name of the mean parameter for this distribution, as well as the link function transforming the mean parameter into the linear predictor.}
\label{table:family_table}
\end{table}

The left panel in Figure \ref{fig:clinton_posterior} shows the marginal posterior distributions of all the coefficients in the model. Panels with the \texttt{party\_id:scale(age)} header are marginal posteriors for the interaction between party affiliation and age, which indicate the slopes for the standardized age predictor in each party affiliation group. These distributions show that, among Democrats, there is not much association between age and voting intentions because the posterior distribution has considerable probability density around zero. However, among both Republicans and Independents, there is a distinct tendency for older respondents to be less likely to indicate an intent to vote for Clinton since their marginal posteriors have most of their density around negative values.

\begin{figure}[htb]
\centering
\includegraphics[width=5in]{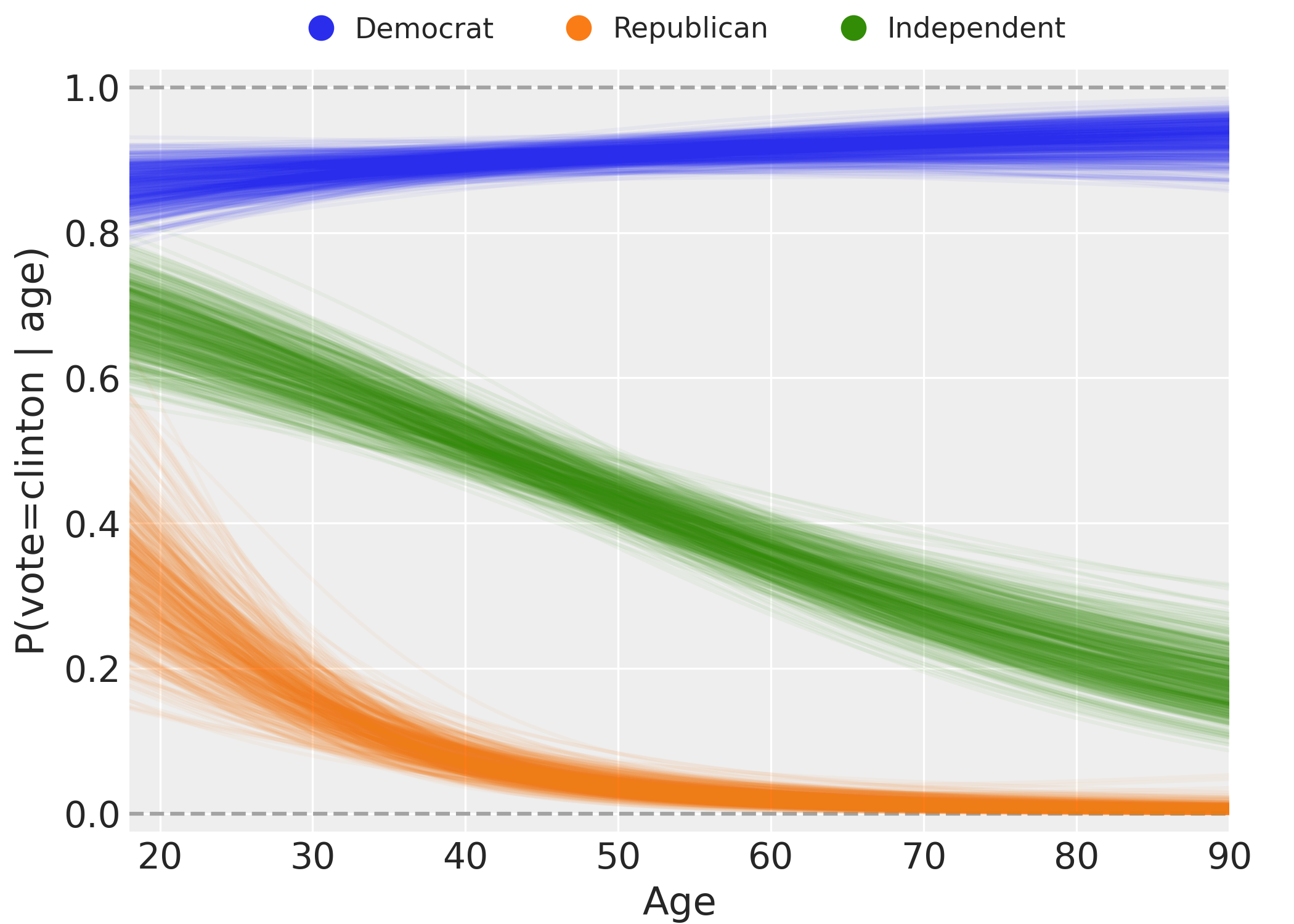}
\caption{Spaghetti plot showing the model predictions and associated uncertainty. The full version of the code that produces this figure can be found in the notebook \texttt{02\_logistic\_regression.ipynb} in the associated GitHub repository.}
\label{fig:clinton_results}
\end{figure}

Key results from the model are summarized in Figure \ref{fig:clinton_results}. This is a \textit{spaghetti plot} showing plausible logistic regression curves for each party identification category, given the data we have now observed. These are obtained by taking parameter values sampled from the posterior and plotting the logistic regression curve implied by those sampled parameters. The spaghetti plot shows the model predictions as well as the uncertainty around those predictions. As we have previously mentioned, for Democrats, the probability of voting for Clinton is not related to age (the probability is almost constant around 0.9 for all ages). However, both older Republicans and older Independents are less likely to vote for Clinton. 

Bambi models have a \texttt{.predict()} method that can be used to obtain both in-sample and out-of-sample predictions. Its only mandatory argument is the \texttt{InferenceData} object returned from the sampling process. This object is modified in-place unless we explicitly add \texttt{inplace=False} in the call. There is also an argument \texttt{kind} indicating the type of prediction, which can be either \texttt{"mean"} (default) or \texttt{"pps"}. The first returns draws from the posterior distribution of the mean, while the latter returns the draws from the posterior predictive distribution. Other possible arguments are \texttt{data} and \texttt{draws}. The former is used to pass a new \texttt{DataFrame} with values for the predictors in order to obtain out-of-sample predictions (if omitted, the original dataset is used), and \texttt{draws} is the number of random draws per chain when using \texttt{kind="pps"}.

In the following code block we use \texttt{model.predict()} to compute the posterior of the mean probability of voting for Clinton for observations in a new dataset that is created to obtain the spaghetti plot in Figure \ref{fig:clinton_results}. Notice we create values for the variable \texttt{age} in the original scale and Bambi automatically handles the transformation for us. Also, since we are not passing any further options, this call computes the posterior of the mean for the new data and modifies \texttt{idata} in-place.

\vspace{10pt}
\begin{minipage}{\linewidth}
\begin{lstlisting}
# Create a new dataset with ages ranging from 18 to 90
# Contains 219 rows
age = np.arange(18, 91)
parties = ["democrat", "republican", "independent"]
new_data = pd.DataFrame({
    "age": np.tile(age, 3),
    "party_id": np.repeat(parties, len(age))
})

# Compute the posterior of the mean P(vote = clinton | age)
# Automatically handles the scaling of 'age'
model.predict(idata, data=new_data)

# Compute mean across chains
posterior_mean = idata.posterior["vote_mean"].values.mean(0)

# Select 500 draws
# Transpose to get an array of shape (219, 500)
posterior_mean = posterior_mean[:500, :].T

# Generate the plot
for i, party in enumerate(["democrat", "republican", "independent"]):
    idx = new_data.index[new_data["party_id"] == party]
    plt.plot(age, posterior_mean[idx], alpha=0.05, color=f"C{i}")
\end{lstlisting}
\end{minipage}
\vspace{10pt}

The call \texttt{model.predict(idata, data=new\_data)} in the previous block of code adds a new variable \texttt{"vote\_mean"} to the \texttt{posterior} group in the \texttt{idata} object. The name of the new variable is the name of the response, plus \texttt{"\_mean"} as a suffix. This variable represents a 3-dimensional object with shape \texttt{(n\_chains, n\_draws, n\_obs)}. These stand for the number of chains and the number of draws per chain we sampled from the posterior, and the number of observations in the new dataset.

\subsection{Hierarchical models}
\label{sec:hierarchical_models}

Bambi makes it easy to fit hierarchical (generalized) linear models\footnote{These types of models are also known as multilevel or mixed effects models.} with common and group specific terms (also known as fixed and random effects, respectively). To illustrate, we conduct a Bayesian reanalysis of the data reported in a registered replication report (RRR) \cite{Wagenmakers2016} of a highly-cited study by \cite{Strack1988}. The original study tested the \textit{facial feedback hypothesis}, which holds that emotional responses are, in part, driven by facial expressions. Strack and colleagues reported in their study that participants rated cartoons as funnier when holding a pen between their teeth (unknowingly inducing a smile) than when holding a pen between their lips (unknowingly inducing a pout). The article has been cited over 1400 times, and has been influential in popularizing the view that emotional experience not only shapes, but can also be shaped by, emotional expression. Yet in a 2016 RRR led by Wagenmakers and colleagues at 17 independent sites \cite{Wagenmakers2016}, spanning over 2500 participants, no evidence in support of the original effect could be found.

We reanalyze and extend the analysis in this RRR using a Bayesian hierarchical model. We fit a hierarchical linear model containing the following terms: (1) the common effect of experimental condition ("smile" vs. "pout"), which is the primary variable of interest; (2) group specific intercepts for the 17 studies; (3) group specific condition slopes for the 17 different study sites; (4) group specific intercepts for all subjects; (5) group specific intercepts for the 4 stimuli used at all sites; and (6) common terms for age and gender (since they are included in the dataset, and could conceivably account for variance in the outcome). Our model departs from the meta-analytic approach used by \cite{Wagenmakers2016} in that the latter allows for study-specific subject and error variances (though in practice, such differences are unlikely to impact the estimate of the experimental condition effect). On the other hand, our approach properly accounts for the fact that the same stimuli were presented in all 17 studies. By explicitly modeling the stimulus as a random variable, we ensure that our inferences can be generalized over a population of stimuli like the ones \cite{Wagenmakers2016} used, rather than applying only to the exact 4 Far Side cartoons that were selected \cite{Judd2012, Westfall2015}.

Bambi allows us to specify priors in two different ways. The first is at model instantiation via the \texttt{priors} argument. We can pass a dictionary where the keys are names of terms in the model, \texttt{"common"} or, \texttt{"group\_specific"} and the values are instances of \texttt{bmb.Prior}. Priors that are passed via the \texttt{"common"} or the \texttt{"group\_specific"} keys are applied to all the common and group specific effects, respectively. The other option is to specify priors after the model is created via the \texttt{.set\_priors()} method. Here we have three arguments: \texttt{priors}, \texttt{common}, and \texttt{group\_specific}. The argument \texttt{priors} expects a dictionary where the keys are term names and the values are instances of \texttt{bmb.Prior}, while \texttt{common} and \texttt{group\_specific} work with \texttt{bmb.Prior} directly.

In the following block of code, we create and fit a model with custom priors using the \texttt{.set\_priors()} method.

\begin{lstlisting}
# Define the model and pass the dataset we want to use.
model = bmb.Model(
    "value ~ condition + age + gender + (1|uid) + (condition|study + stimulus)"
    data=long, dropna=True
)

# Set a custom prior on common predictors
common_prior = bmb.Prior("Normal", mu=0, sigma=0.5)
model.set_priors(common=common_prior)

# Set a custom prior on group specific factor standard deviations
group_specific_sd = bmb.Prior("HalfNormal", sigma=1)
group_specific_prior = bmb.Prior("Normal", mu=0, sigma=group_specific_sd)
model.set_priors(group_specific=group_specific_prior)

# Set a custom prior on Intercept and age
priors = {
    "Intercept": bmb.Prior("Normal", mu=0, sigma=3),
    "age": bmb.Prior("Normal", mu=0, sigma=0.3)
}
model.set_priors(priors=priors)

# Increase taget_accept to 0.95.
idata = model.fit(target_accept=0.95)
\end{lstlisting}

We have omitted the steps necessary to obtain the \texttt{long} dataset, as these are unrelated to Bambi. These steps, together with all the code to reproduce this example, are included in the notebook \texttt{03\_hierarchical\_model.ipynb}, which is in the GitHub repository of this paper.

There are some novel features in the code block above that are worth mentioning. Notice the \texttt{(condition|study + stimulus)} term in the model formula. This is a shorthand that leverages the distributive property of the \texttt{|} operator and is equivalent to have written both \texttt{condition|study} and \texttt{condition|stimulus}. We also used \texttt{dropna=True} to tell Bambi to discard any rows with missing values in any of the columns in the model. This automatically dropped 33 out of 6940 rows from the dataset.

Next, we specified custom priors. Specifically, we first indicated that all common terms have a \texttt{Normal} prior with mean 0 and standard deviation of 0.5. Then, we indicated the variances of all group specific terms to be modeled using a \texttt{HalfNormal} distribution with \texttt{sigma=1}. And finally, we specified priors for the intercept and the age terms. These are given by \texttt{Normal} distributions centered at 0 with standard deviations equal to 3 and 0.3, respectively. These non-default prior has no discernible impact on the posterior because the dataset is relatively large, but results in an improved sampling performance compared to using the default priors. As the package documentation explains, one can easily specify a completely different prior for each model term, and any one of the many preexisting distributions implemented in PyMC can be assigned. Finally, we have increased \texttt{target\_accept} from the default \texttt{0.8} to \texttt{0.95}. This is a parameter that is passed to NUTS, the sampler, so the step size is tuned such that it approximates this acceptance rate. Higher values often work better for problematic posteriors.

\begin{figure}[H]
\centering
\includegraphics[width=5.5in, height=4.6in]{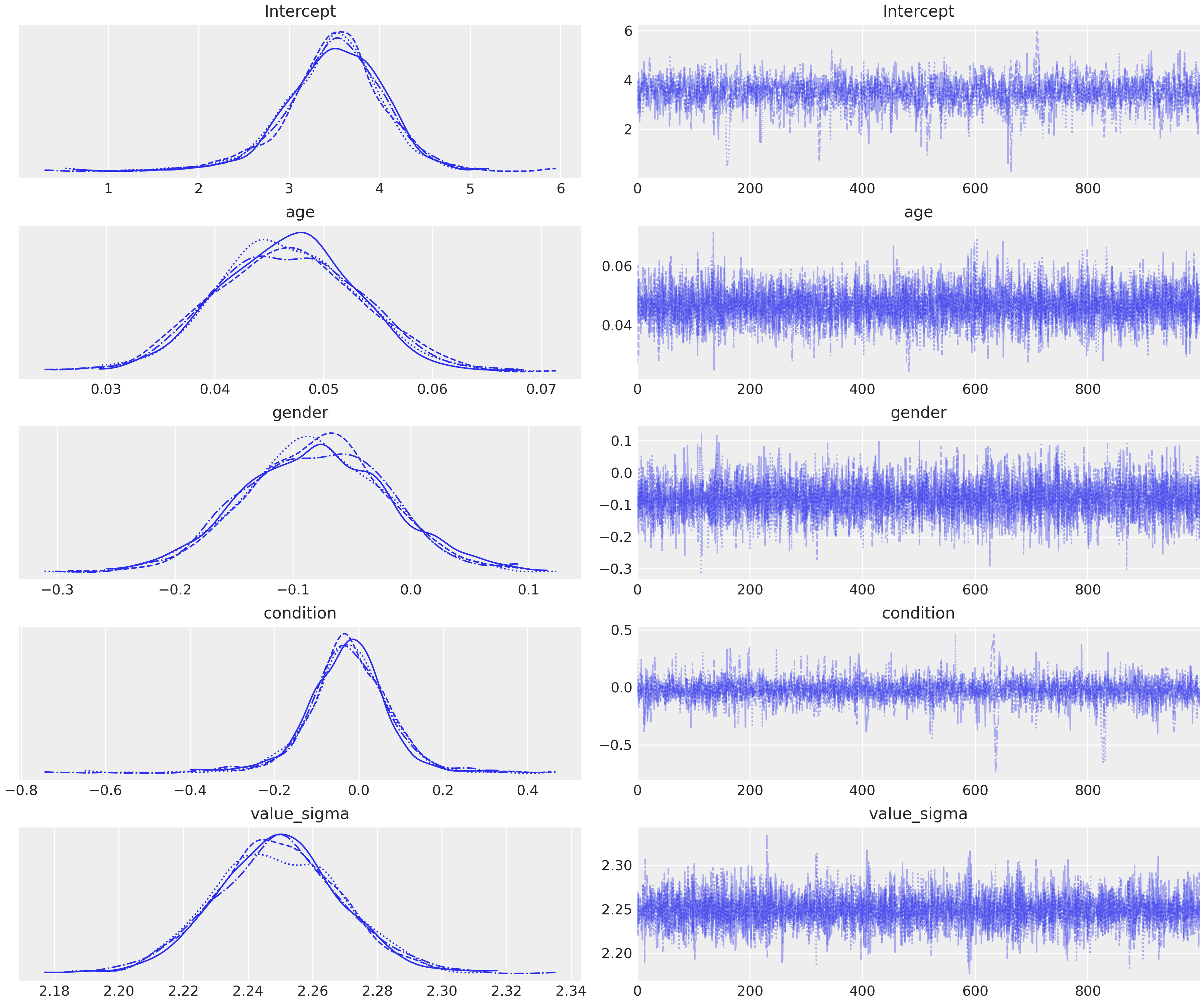}
\caption[]{Marginal posterior distributions and sample traces (right) of parameter estimates for common terms. The parameter labeled \texttt{value\_sigma} is the residual standard error, usually denoted $\sigma$.}
\label{fig:rrr_posterior_common}
\end{figure}

\begin{figure}[H]
\centering
\includegraphics[width=5.5in, height=8.25in]{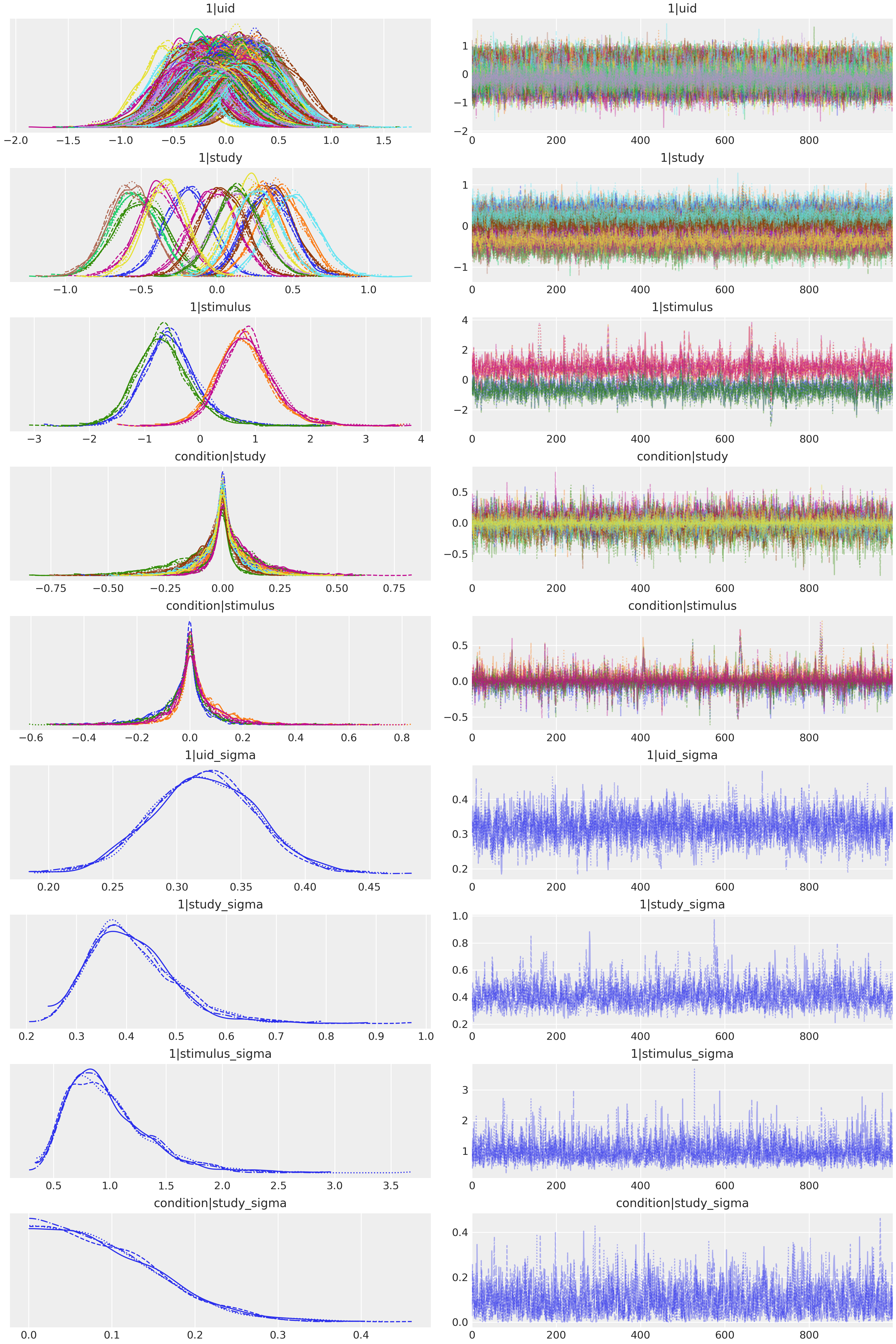}
\caption[]{Marginal posterior distributions and sample traces of parameter estimates for all group specific terms. The terms with the suffix \texttt{\_sigma} are the standard deviations of group specific terms (e.g., \texttt{1|uid\_sigma} is the SD of the means of the individual subject intercepts).}
\label{fig:rrr_posterior_group_specific}
\end{figure}

Inspection of the results from Figures \ref{fig:rrr_posterior_common} and 
\ref{fig:rrr_posterior_group_specific} reveals essentially no effect of the experimental manipulation, consistent with the findings reported in \cite{Wagenmakers2016}, including the observation that the variation across sites is surprisingly small in terms of both the group specific intercepts (\texttt{1|study}) and the group specific slopes (\texttt{condition|study}). One implication of this observation is that the constitution of the sample, the language of instruction, or any of the hundreds of other potential between-site differences, appear to make much less of a difference to participants' comic ratings than one might have intuitively supposed. Interestingly, our model also highlights an additional point of interest not discernible from the results reported by \cite{Wagenmakers2016}: the posteriors for \texttt{1|stimulus} are much wider than the posteriors for the other factors, which means the stimulus level variance is very large compared to the others. This is problematic, because it suggests that any effects one identifies using a conventional analysis that fails to model stimulus effects could potentially be driven by idiosyncratic differences in the selected comics. Note that this is a problem that affects both the RRR and the original Strack study equally. In other words, regardless of whether the RRR would have obtained a statistically significant replication of the Strack study given different stimuli, if the effect is strongly driven by idiosyncratic properties of the specific stimuli used in the experiment—which is not unlikely, given that the results are based on just four stimuli drawn from a stimulus population—that is likely quite heterogeneous, then there would have been little basis for drawing strong conclusions from that result in the first place. Either way, the moral of the story is that any factor that can be viewed as a sample from some population that one intends to generalize one's conclusions over (such as the population of funny comics) should be explicitly included in one's model \cite{Judd2012, Westfall2015}. Bambi makes it very easy to fit such models within a Bayesian framework. 

\section{Default prior choice}
\label{sec:prior_choice}

The goal of the default prior implementation in Bambi is to automatically provide prior distributions that are weakly informative for a wide range of use cases. Such weakly informative priors aim to provide slight regularization and help stabilize the computation, rather than incorporating prior domain knowledge into the model. These priors work well in many applications, however it is not possible to guarantee that they will be suitable for every scenario. Thus, users are advised to follow good practices, including the use of convergence diagnostics and model checking \cite{BayesianWorkflow2020, martin2021}.

Bambi obtains default priors using a procedure very similar to the one in the rstanarm library, whose online documentation can be found in \url{https://mc-stan.org/rstanarm/articles/priors.html}.

\subsection{Regression coefficients for common effects}

For all models and families, we use independent Normal priors centered at 0 for the common effects regression coefficients. Their standard deviation depends on the model family, and the scale of the response and the predictor.

Suppose we have a response $Y$ and a predictor $X_k$. Then, the default prior for the regression coefficient $\beta_k$ is

\begin{equation}
\beta_k \sim \text{Normal} \left (0, 2.5 \cdot \frac{\text{sd}(Y)}{\text{sd}(X_k)}\right)
\end{equation}

where $\text{sd}(X_k)$ is the sample standard deviation of the predictor $X_k$ and $\text{sd}(Y)$ is the sample standard deviation of the response $Y$ if the model family is either \texttt{"gaussian"} or \texttt{"t"}, and 1 otherwise.

\subsection{Intercept}

The default prior for the intercept $\beta_0$ follows a different scheme for two reasons. First, it is represented as a constant predictor of ones in the design matrix. And second, Bambi centers the predictors when the model contains an intercept, which means the prior for the intercept actually refers to the prior of the centered intercept.

We first note that in Ordinary Least Squares (OLS) regression we have $\beta_0 = \bar{Y} - \beta_1\bar{X}_1 - \beta_2\bar{X}_2 - \dots$ where $\bar{Y}$ represents the mean of $Y$, so we can set the mean of the prior on $\beta_0$ to

$$
\text{E}[\beta_0] = \bar{Y} - \text{E}[\beta_1]\bar{X}_1 - \text{E}[\beta_2]\bar{X}_2 - \dots
$$

In practice, both the priors on the slopes and the centered predictors will have a zero mean, so the mean of the prior on $\beta_0$ will typically reduce to $\bar{Y}$.

Now for the variance, and assuming independence of the slope priors, we have:

\begin{equation}
    \text{var}(\beta_0) = \frac{\text{var}(Y)}{n} + \bar{X}_1^2\text{var}(\beta_1) + \bar{X}_2^2\text{var}(\beta_2) + \dots
\label{b0var}
\end{equation}

In other words, once we have defined the priors on the slopes, we can combine this with the means of the predictors to find the implied variance of $\beta_0$. Our default prior for intercepts is a Normal distribution with mean and variance defined as above, except that the $\frac{\text{var}(Y)}{n}$ term in the Equation \ref{b0var} is replaced by $\text{var}(Y)$, so that the intercept prior will not be too narrow when the predictors are centered and the sample size is large.

\subsection{Auxiliary parameters}

Some likelihood functions in the built-in families have other parameters than the mean $\mu$. Some examples are the standard deviation in a Normal likelihood or the shape parameter in a Gamma. These parameters are not modeled by a transformation of the linear predictor, and we thus put a prior on them. Table \ref{table:family_priors} summarizes the auxiliary parameters for these families and their default priors.

\begin{table}[t!]
\centering
\begin{tabular}{lllp{7.4cm}}
\hline
Family & Auxiliary parameter & Default prior \\  \hline
beta & $\kappa$ & HalfCauchy($\beta = 1$) \\
gamma & $\alpha$ & HalfCauchy($\beta = 1$) \\
gaussian & $\sigma$ & HalfStudentT($\nu=4$, $\sigma=\text{sd}(y)$) \\
negativebinomial & $\alpha$ & HalfCauchy($\beta = 1$) \\
t & \makecell[l]{$\sigma$ \\ $\nu$} & \makecell[l]{HalfStudentT($\nu=4$, $\sigma=\text{sd}(y)$) \\ Gamma($\alpha=2$, $\beta=0.1$)} \\
wald & $\lambda$ & HalfCauchy($\beta = 1$)
\end{tabular}
\caption{Built-in families, auxiliary parameters, and their default priors. The meaning of these parameters is the one given in the PyMC library.}
\label{table:family_priors}
\end{table}

\subsection{Group specific effects}

As is customary with mixed models, group specific (random) effects are assumed to be Normally distributed. The default prior variances of those Normal distributions are based on the idea that, generally speaking, the greater the prior variance is of the corresponding common effect coefficient, the greater should be the prior variance of the group specific effect variance. We implement this idea by using Half-Normal distributions for the group specific effect standard deviations, each with parameter $\sigma$ set equal to the prior standard deviation of the corresponding common effect. If the common part of the model does not include the corresponding common effect, then we consider an augmented model in which the common part of the model \textit{does} include the corresponding common effect, and we compute what would be the mean and standard deviation of the prior for this common effect using the methods described previously, and then set $\sigma$ equal to this implied prior standard deviation.

\subsection{Previous default priors}

Bambi originally implemented an algorithm that would set default priors on the scale of the implied partial correlation between the predictors and the response that is described in \cite{Westfall2017}.  

One of the major drawbacks of this method is that it depends on maximum likelihood estimates (MLE) for the parameters in the model, meaning it is not available whenever the MLE does not exist, as it is in the case of complete separation or when there are more columns than observations in the design matrix. In addition, the implementation uses the \texttt{GLM} module from the statsmodels library to compute maximum likelihood estimates. Thus, the range of models that were implemented in Bambi was limited by the models available in statsmodels which lacks families such as the \texttt{"beta"} or \texttt{"t"} that are now part of Bambi.

At the moment of writing, this default prior algorithm is still available, but it is not the default choice anymore and will be removed in the future. For now, users can use this method by setting \texttt{automatic\_priors="mle"} when instantiating a \texttt{Model}.

\subsection{Limitations and future extensions}

Our default prior system is based on independent Normal priors for all slopes (either common or group specific), so that their joint prior distribution is multivariate Normal with a diagonal covariance matrix. It is possible that allowing this multivariate Normal to have non-zero covariances would make sense. We have recently added a new feature that allows group specific coefficients to have multivariate Normal prior whose covariance matrix has a LKJ prior. This approach is inspired on the prior for the group specific coefficients in  rstanarm \cite{Goodrich2020}.

Even though our default priors work well in practice, we have performed simulations with fake data to explore the weaknesses of our proposal. The results suggest that there can be cases where the resulting priors are too narrow, and consequently the regularization is inadvertently greater than one might expect. This could also pose a problem for the sampler, which is often manifested as a high number of divergences. 
 
For example, if in a regression setting the variability of $X_j$ is much higher than the variability of $Y$ and $|\beta_j| \gg 0$, we may end up with a prior on $\beta_j$ that puts very little probability around its true value. 

We also believe the priors for auxiliary parameters could be further improved. For example, the peak of the HalfStudentT prior for $\sigma$ in the \texttt{"gaussian"} and \texttt{"t"} families is at 0, which implies a coefficient of determination $R^2=1$. On the other hand, this prior has a heavy tail, so the bulk of the distribution is away from zero. In a future work, we could compare the current approach with one that uses a right skewed distribution whose peak is not at 0 and see which one tends to work best in most situations.

\section{Formula specification}
\label{sec:formula}

A model formula is a string of the form \texttt{"resp ~ expr"}, where \texttt{resp} indicates the response variable and \texttt{expr} is an expression that determines the design matrices $\mathbf{X}$ and $\mathbf{Z}$ for the common and group specific effects, respectively. Bambi uses formulae \cite{bambinoscapretto2021}, an implementation of model formulas written by Bambi developers. 

\subsection{Formulae}

formulae is very similar to the model formula implementation in R in both its syntax and semantics. Most formulas that work in R are expected to work in formulae similarly as long as you write Python code instead of R when including function calls. 

\subsubsection{Available operators}

A list of available operators together with their description can be found in Table \ref{table:formulae_operators}. There, operators are sorted from highest to lowest precedence. Operators in the same section, delimited by a horizontal line, have the same precedence level. Also, note that formula expressions are interpreted from left to right, but as may be naturally expected, expressions within parenthesis are resolved first, and then they can be used to override precedence rules.

\begin{table}[t!]
\centering
\begin{tabular}{clc}
\hline
Op. & Description \\ \hline

    \texttt{**} &
    \makecell[l]{
        Power operator. It takes a set of terms on the left, an integer $n$ \\ 
        on the right, and returns all the interaction between the terms \\
        up to order $n$.
    } \\ \hline
    \texttt{:} & 
    Interaction between operands. \\ \hline
    \texttt{*} & 
    \makecell[l]{
        Full interaction. Includes the interaction between operands as\\   
        well as the operands themselves. \\ 
        \texttt{a*b} is a shorthand for \texttt{a + b + a:b}.
    } \\
    \texttt{/} & 
    \makecell[l]{
        \texttt{a / b} is a shorthand for \texttt{a + a:b}. It is rightward distributive but\\
        not leftward distributive over \texttt{+}. \texttt{a / (b + c)} is equal to \\ 
        \texttt{a + a:b + a:c} but \texttt{(a + b)/c} is equal to \texttt{a + b + a:b:c}.
    } \\ \hline
    \texttt{+} & 
    \makecell[l]{
        Computes a set union between terms on the left and terms on\\
        the right. This means that \texttt{a+a} is \texttt{a}.
    } \\
    \texttt{-} & 
    \makecell[l]{
        Computes a set difference between terms on the left and terms on\\
        the right. But since we parse from left to right, \texttt{x + y - x}\\
        is \texttt{y} but \texttt{y - x + x} is equal to \texttt{y + x}.
    } \\ \hline
    \texttt{|} & 
    \makecell[l]{
        Interaction-like operator that indicates a group specific effect.\\
        The expression on the left-hand side contains an implicit\\
        intercept. The right-hand side is interpreted as a categorical grouping variable. 
    } \\ \hline
    \textasciitilde & 
    \makecell[l]{
        Separates the left-hand side and right-hand side of a formula.\\ 
        The left-hand side  represents the response while the right-hand \\ 
        side is an expression that determines the design matrices $X$ and $Z$.} \\ \hline
\end{tabular}
\caption{Formulae (and Bambi) built-in operators}
\label{table:formulae_operators}
\end{table}

\subsubsection{Group specific effects}

Group specific effects are specified and interpreted by formulae in the same way as in the R package lme4, with the exception that formulae lacks the \texttt{||} operator to specify uncorrelated group specific intercept and slope. That is, group specific effects are of the form \texttt{(expr|factor)}\footnote{Parentheses are optional, but we almost always use them because the pipe operator has lower precedence than all the operators, except for the tilde \textasciitilde.}. The expression \texttt{expr} is evaluated as a model formula itself, producing a design matrix following the same rules than those for common effects, and \texttt{factor} is interpreted as a grouping variable. Then, the computation of the group specific effects matrix, $Z$, is carried out exactly as specified in Section 2.3 of \cite{lme4}.

\subsubsection{Differences with R and Patsy}

Earlier versions of Bambi relied on Patsy \cite{Smith2018} to parse model formulas and construct design matrices. But its lack of built-in support for mixed effects made it cumbersome to specify group level effects, which had to be passed as a separate list. For example, to fit a regression of \texttt{y} on \texttt{x}, with each group in \texttt{g} having a group specific intercept and slope, the user would write

\begin{lstlisting}[numbers=none]
model.fit("y ~ x", group_specific=["x|g"])
\end{lstlisting}

Instead, formulae enables the specification of group level effects via the \texttt{|} operator within a single model formula. Then, the previous method call is simplified to 

\begin{lstlisting}[numbers=none]
model.fit("y ~ x + (x|g)")
\end{lstlisting}

The main difference between formulae and R relies on how we encode categorical variables when constructing design matrices. For some specifications that include categorical variables R produces over- or under-specified model matrices. To avoid that, formulae uses an algorithm introduced in Patsy to decide whether to use a full- or reduced-rank coding for categorical variables that ensures it always returns model matrices that are full-rank. But in contrast to Patsy, only a reference term encoding is available in formulae for now.

formulae also includes some syntactic sugar to improve the user experience. For example, to disambiguate an operation like a sum between two terms, R requires wrapping the sum with the \texttt{I()} function call, such as in \texttt{I(x + y)}. In formulae it can be simplified to \texttt{\{x + y\}}, but the R version still works. Also, while non-syntactic names like \texttt{My question?} have to be wrapped with \texttt{Q()} in Patsy, they can be escaped with the backtick operator in formulae such as in \texttt{"response ~ \textasciigrave My question?\textasciigrave"}.

formulae is not Patsy nor R, but our attempt to take the best of R and Patsy and make it available for Bambi. The result is an implementation of the formula language where you can specify common and group specific effects in a single string, you have an algorithm to build model matrices that ensures the result is structurally full rank, and we as Bambi developers have the possibility to modify the source code to incorporate (or modify) features in order to make it easier to work with Bayesian GLMMs.

\section{Discussion and conclusions}
\label{discussion}

We have introduced a high-level Bayesian model building interface that combines a formula notation similar to those found in the popular R packages lme4 and brms with the flexibility and power of the stat-of-the-art probabilistic programming framework PyMC. The example applications presented here illustrate how Bambi makes it possible to fit sophisticated Bayesian generalized linear multilevel models with little programming knowledge, using a formula-based syntax that will already be familiar to many users, or that is easy to learn for those unfamiliar with it.

The Bayesian approach to fitting generalized linear multilevel models is attractive for several reasons. Practically speaking, the ability to inject outside information into the analysis in a principled way, in the form of the prior distributions of the parameters, can have a beneficial regularizing effect on parameter estimates that are computationally difficult to pin down. For example, the variances of group effect terms can often be quite hard to estimate precisely, especially for small datasets and unbalanced designs. In the traditional maximum likelihood setting, these difficulties often manifest in point estimates of the variances and covariances that defy common sense, or in outright failure of the model to successfully converge to a satisfactory solution. Setting an informative Bayesian prior on these variance estimates can help bring the model estimates back down to earth, resulting in a convergent fitted model that is much more believable \cite{Chung2015}. As a second example, it is well known that categorical outcome models such as logistic regression can suffer problems due to quasi-complete separation of the outcome with respect to the predictors included in the model, which has the effect of driving the parameter estimates toward unrealistic values approaching minus or plus infinite. These problems are further exacerbated in mixed-effects logistic regression, where separation in some individual clusters can easily distort the overall common parameter estimates, even though there is no separation at the level of the whole dataset. Again, informative priors can help to rein in these diverging parameter estimates \cite{Gelman2008}.

For users with more programming experience, the probabilistic programming paradigm that PyMC supports may also confer additional benefits. A key benefit of this approach is that researchers can theoretically fit any model they can write down; no analytical derivation of a solution is required. Because Bambi is simply a high-level interface to PyMC, in practice, researchers can use Bambi to very quickly specify a model, and subsequently elaborate on, or extend that model, in native Python code. Subject to limitations of the state-of-the art Bayesian inference methods, users can in principle fit any model that allows variables to be modeled as probability distributions, including arbitrary non-linear transformations.

When working with Bayesian models, there are a series of related tasks that need to be addressed besides inference itself \cite{Kumar2019, BayesianWorkflow2020, martin2021}. These tasks include diagnosing the quality of the inference, model criticism and model comparison. While Bambi is an interface for model building and inference, its tight integration with ArviZ helps users to perform this non-inferential tasks in a more fluid manner.

Importantly, the open source nature of Bambi, and in particular, its reliance on numerical packages that already have well-established user bases and developer communities, means that the available functionality will continue to grow rapidly. Our hope is that many researchers accustomed to frequentist methods will find Bambi sufficiently intuitive and familiar to warrant adopting a Bayesian approach for at least some classes of common analysis problems. In the future, we would like to improve Bambi's default priors, which is a topic that will benefit from more research. On a more immediate future, we would like to include features that are currently missing, such as support for splines or Gaussian processes. 

\section{Acknowledgement}
The work was supported by the National Agency of Scientific and Technological promotion, ANPCyT (Grant No PICT-0218). We acknowledge the computational resources provided by the Aalto Science-IT project and support by the Academy of Finland Flagship program: Finnish Center for Artificial Intelligence, FCAI, and the Technology Industries of Finland Centennial Foundation (grant 70007503; Artificial Intelligence for Research and Development).

\bibliography{biblio}

\newpage
\appendix
\section{Appendix}
\subsection{Multiple linear regression}
\label{sec:multiple_regression_appendix}

In the context of Ordinary Least Squares (OLS) regression, it is possible to convert the regression coefficient of a given predictor into a partial correlation by multiplying it by a constant. Given an outcome $Y$, a predictor $X_k$, and a set of predictors $\mathbf{X}_{-k}$ not containing $X_k$, one can convert the estimated slope for $X_j$ to a partial correlation between $X_k$ and $Y$ controlling for $\mathbf{X}_{-k}$ using the following identity

\begin{equation} \label{a1eq1}
    \rho_{X_j Y\cdot \mathbf{X}_{-k}} = 
    \beta_k \frac{\text{sd}(X_k)}{\text{sd}(Y)}
    \sqrt{\frac{1-R_{X_k \mathbf{X}_{-k}}^2}{1-R_{Y \mathbf{X}_{-k}}^2}}
\end{equation}

where $\beta_k$ is the estimated slope for $X_k$, $R_{X_k \mathbf{X}_{-k}}^2$ is the $R^2$ from a regression of $X_k$ on $\mathbf{X}_{-k}$, and $R_{Y \mathbf{X}_{-k}}^2$ is the $R^2$ from a regression of $Y$ on $\mathbf{X}_{-k}$.

For the calculations in this section, we used the \texttt{posterior} object contained in \texttt{idata} \texttt{InferenceData} object. Also, we require that pandas and statsmodels API are loaded as \texttt{pd} and \texttt{sm}, respectively. We first compute the needed statistics, and then we convert the slopes to partial correlations.

\begin{lstlisting}
# Select all terms but the Intercept
terms = [t for t in model.common_terms.values() if t.name != "Intercept"]

# Common effects design matrix (excluding intercept/constant term)
x_matrix = [pd.DataFrame(x.data, columns=x.levels) for x in terms]
x_matrix = pd.concat(x_matrix, axis=1)

samples = idata.posterior
varnames = ["o", "c", "e", "a", "n"]

# Compute the needed statistics
sd_x = x_matrix.std()
sd_y = data["drugs"].std()
r2_x = pd.Series(
    {
        x: sm.OLS(
            endog=x_matrix[x],
            exog=sm.add_constant(x_matrix.drop(x, axis=1))
        )
        .fit()
        .rsquared
        for x in list(x_matrix.columns)
    }
)
r2_y = pd.Series(
    [
        sm.OLS(
            endog=data["drugs"],
            exog=sm.add_constant(data[[p for p in varnames if p != x]]),
        )
        .fit()
        .rsquared
        for x in varnames
    ],
    index=varnames
)

constant = (sd_x[varnames] / sd_y) * ((1 - r2_x[varnames]) / (1 - r2_y))**0.5
pcorr_samples = samples[varnames] * constant 
\end{lstlisting}

Notice, however, this practical approach is not free of caveats. As we mentioned in the first paragraph of this section, formula \ref{a1eq1} is valid in the context of OLS regression, where the $\beta_k$ and $R^2$ values are considered fixed quantities. In our case, we are plugging $R^2$ values obtained from a OLS regression fit together with posterior draws to obtain an approximation to the posterior distribution of the partial correlations.

\end{document}